\documentclass[submission, Phys]{SciPost}
\usepackage[english]{babel}
\usepackage{amsthm}
\usepackage{titlesec}
\titlespacing*{\section}{0pt}{0.07\baselineskip}{0.07\baselineskip}
\theoremstyle{definition}
\newtheorem{definition}{Definition}[section]

\theoremstyle{remark}

\usepackage{comment}

\begin{document}

\begin{center}{\Large \textbf{
Necessity of Quantizable Geometry for Quantum Gravity
}}\end{center}

% TODO: write the author list here. Use initials + surname format.
% Separate subsequent authors by a comma, omit comma at the end of the list.
% Mark the corresponding author with a superscript *.
\begin{center}
A. K. Mehta\textsuperscript{1}
\end{center}

\begin{center}
{\bf 1}Asia Pacific Centre for Theoretical Physics, Pohang, Republic of Korea
\\
*abhishek.mehta@apctp.org
\end{center}

\begin{center}
\today
\end{center}
% For convenience during refereeing: line numbers
%\linenumbers

\section*{Abstract}
{\bf
In this paper, Dirac Quantization of $3D$ gravity in the first-order formalism is attempted where instead of quantizing the connection and triad fields, the connection and the triad 1-forms themselves are quantized. The exterior derivative operator on the space of differential forms is treated as the `time' derivative to compute the momenta conjugate to these 1-forms. This manner of quantization allows one to compute the transition amplitude in $3D$ gravity which has a close, but not exact, match with the transition amplitude computed via LQG techniques. This inconsistency is interpreted as being due to the non-quantizable nature of differential geometry.
% TODO: write your abstract here.

}

\vspace{10pt}
\noindent\rule{\textwidth}{1pt}
\tableofcontents\thispagestyle{fancy}
\noindent\rule{\textwidth}{1pt}
\vspace{10pt}
\section{Introduction and Summary of Results}
Quantization of gravity is one of the most outstanding problems in theoretical physics. Over the years, various approaches have been developed to tackle this problem. One approach involves foliating the manifold with spacelike hypersurfaces and quantizing the induced metric field $h_{\mu\nu}$. This method of quantization of gravity involves a significant computational challenge. This challenge is compounded by presence of gauge symmetry and topology where gauge fixing requires special attention and topology leads to additional complications in the spacetime evolution\cite{anderson1997global}. Since, this approach fixes the underlying geometry of the spacetime up to the field variable $h_{\mu\nu}$, the geometry at the quantum level is simply fluctuations of the metric field or gravitons\cite{Wiltshire:1995vk}. In approaches like string theory, one gets around the tedious problem of metric field quantization by developing a quantum theory of gravity by considering string propagation in a target spacetime which reproduces General Relativity (GR) as a low-level effective theory\cite{polchinski2005string}. The geometry at the quantum level is, therefore, characterized by string fluctuations. Other approaches too have been proposed that ``deal'' with gravity rather than quantizing it\cite{verlinde2011origin, padmanabhan2016exploring}.\\\\
Considering the overwhelming difficulty and tediousness of formulating a quantum gravity, one may wonder if a quantum theory of gravity even exists, and indeed, its existence has been called into question quite recently and correspondingly, alternative formulations have been proposed where gravity is treated as purely classical whilst the particles coupling to gravity is quantized\cite{oppenheim2018post,oppenheim2022constraints}. However, before we go on to quantize gravity (or perhaps, reject it's possibility), we should note that gravity or GR is mathematically a theory of differential manifolds. Therefore, naively, quantizing gravity at some point must involve ``quantizing'' differential geometry itself. However, no such possibility has been explored so far in literature. Although heuristically speaking, approaches like Loop Quantum Gravity (LQG) does come quite close to doing so, however, in practice it is the triangulations of the manifold that is quantized rather than the original manifold itself\cite{rovelli2015covariant}. The quantization procedure is independent of the triangulations and the usual differential geometry emerges an asymptotic limit of the quantum theory. Also, LQG was developed to mirror the consequences of the Ponzano-Regge model, a model of discretized GR due to which a physical consequence of LQG is that the quantum spacetime geometry must be discrete. More concretely, the geometry at the quantum level is characterized by a spinfoam.\cite{rovelli2004quantum}\\\\
Still, however, an approach that takes the idea of ``quantizing'' differential manifolds and their associated geometry in its \emph{literal} sense has not yet been developed so far. Therefore, this paper considers this direct possibility by treating the differential forms as ``fields'' and by attempting to ``quantize'' them via the Dirac procedure. This is as literal as possible one can get to ``quantizing'' differential geometry. In this paper, we work with $3D$ gravity and treat the connection 1-form and the triad 1-form as fields with ``grassmannian'' statistics and perform a Dirac Quantization of these objects. An initial quantization attempt in the first-order formalism doesn't yield anything fruitful. However, a simple modification of the geometry eventually reproduces the Ponzano-Regge amplitude in LQG closely, yet not exactly. This can be interpreted as a demonstration of the non-quantizable nature of differentiable manifolds and, therefore, establishing a requirement of a ``quantizable geometry'' to describe quantum gravity. Finally, the subtle differences between the transition amplitude computed through LQG and a direct quantization of the geometry is discussed.

\section{$3D$ gravity}
The Einstein-Hilbert (EH) action in first-order formalism is given by
\begin{align}
    S  &= \frac{1}{2\pi G_N}\int_{\mathcal{M}}e_A \wedge F_{BC}~ \epsilon^{ABC} \notag\\
    &= \frac{1}{2\pi G_N}\int_{\mathcal{M}}e_A \wedge F^{A} \label{EH3D}
\end{align}
where
\begin{align}
    F^{A} = d\omega^{A}+\frac{1}{2}\epsilon^{ABC}\omega_{B}\wedge\omega_{C}
\end{align}
where $\omega^A = \epsilon^{ABC}\omega_{BC}$ is the connection 1-form and $e_A$ is the triad 1-form on the pseudo-Riemannian manifold $\mathcal{M}$. We will now perform a quantization of these 1-forms. Since, on the space of forms $\oplus_r\Omega^r(\mathcal{M})$, the only meaningful derivative is the exterior derivative $d$, therefore, the exterior derivative should be interpreted as the `time' derivative on the space of forms. We can show that this interpretation consistently reproduces the Euler-Lagrange equations. For example, consider the scalar action given by
\begin{align}
    S = \int_{\mathcal{M}} (d\phi*d\phi + m^2\phi*\phi) \equiv \int_{\mathcal{M}}\mathcal{L}
\end{align}
The Euler-Lagrange equation where the exterior derivative is the `time' derivative then looks like
\begin{align}
    &d\left(\frac{\delta\mathcal{L}}{\delta(d\phi)}\right) -\frac{\delta\mathcal{L}} {\delta\phi}=0\\
    &\implies d*d\phi -m^2*\phi = 0 \implies *d*d\phi-m^2\phi = 0
\end{align}
which is precisely the Klein-Gordon equation in a curved space. Therefore, sticking with this interpretation, the conjugate momenta for \ref{EH3D} are as follows 
\begin{align}
    &\pi_{A}^{\omega} = \frac{\delta S}{\delta (d\omega^{A})} =\frac{e_A}{2\pi G_N}%+2\beta\star F
    \\
    & \pi^e_A = \frac{\delta S}{\delta (de^A)} = 0\label{APB}
\end{align}
where we have the following anticommuting Poisson brackets
\begin{align}
    \{\pi^{\omega}_A, \omega^{B}\} = \delta_A^{~B} \quad \{\pi^e_{A}, e^B\} = \delta_A^B 
\end{align}
The anticommuting Poisson brackets is used due to the ``fermion'' statistics of 1-form $$\eta_1\wedge\eta_2 = -\eta_2\wedge\eta_1 \quad \eta_1, \eta_2 \in \Omega^1(\mathcal{M})$$
From the above, we can identify the following primary constraints
\begin{align}
    &\chi_{1A}\equiv\pi_{A}^{\omega} - \frac{e_A}{2\pi G_N}\approx 0 \\
    &\chi_{2A}\equiv\pi_{A}^{e} \approx 0 
    \label{pc}
\end{align}
The constraints are second class. Then the preliminary Hamiltonian density is as follows
\begin{align}
    H_1 =-\frac{1}{2}\epsilon^{ABC}\pi^{\omega}_A\wedge\omega_{B}\wedge \omega_{C}+ c^A_1\wedge\chi_{1A} + c^A_2\wedge\chi_{2A}
\end{align}
The Dirac matrix due to the constraints are as follows
\begin{align}
    \mathcal{C}_{ij} = \{\varphi_{i}, \varphi_{j}\}= \begin{pmatrix}
         0 & -\frac{I}{2\pi G_N} \\
        -\frac{I}{2\pi G_N} & 0\\
    \end{pmatrix}_{ij}\label{DM1}
\end{align}
where $\varphi_{i} \equiv (\chi_{1A}, \chi_{2A})$. The inverse of which is given by
\begin{align}
    \mathcal{C}^{-1}_{ij} = \begin{pmatrix}
       0 & -2\pi G_NI \\
        -2\pi G_N I & 0\\
    \end{pmatrix}_{ij}
\end{align}
Using the above the Dirac Hamiltonian may be computed to give
\begin{align}
    H =-\frac{1}{2}\epsilon^{ABC}\pi^{\omega}_A\wedge\omega_{B}\wedge \omega_{C}
\end{align}
with the following Dirac bracket
\begin{align}
    \{\pi^{\omega}_A, \omega^{B}\}_D = \delta_A^{~B}
\end{align}
We now elevate it to the quantum oscillator algebra
\begin{align}
    [\pi^{\omega}_A, \omega^{B}]_{+} = i \delta_A^{~B} \quad [\omega^A, e^B]_{+} = 2\pi G_N \delta^{AB}
\end{align}
We now define
\begin{align}
    J^A = \epsilon^{ABC}\pi^{\omega}_B\wedge\omega_C 
\end{align}
such that we have\footnote{`$+$'- anticommutator, `$-$' - commutator}
\begin{align}
    [J^A, J^B]_{-} = i\epsilon^{ABC}J^C \quad [\omega^A, J^B]_{-} = i\epsilon^{ABC}\omega_C \quad [\omega_A, \omega_B]_{+} = 0\label{Aljeb}
\end{align}
which is very reminiscent of the $U$ and $L$ variable commutation relation in LQG. Notice how the quantized differential 1-forms behave exactly like Grassmannian operators. Therefore, one can heuristically argue that quantization of geometry is characterized by the following map
\begin{align}
    \Omega^1(\mathcal{M}) \overset{Dirac}{\rightarrow} \{\theta\}
\end{align}
where $\{\theta\}$ represents the set of Grassmann numbers. In terms of these variables, the Dirac Hamiltonian becomes
\begin{align}
    H =-\frac{1}{2} \omega_A \wedge J^A
\end{align}
Now, notice that in this Hamiltonian $\omega_A$ is very much like the axis of rotation while $J^A$ is the generator of $SU(2)$ group as can be inferred from the commutation algebra in \ref{Aljeb}. Therefore, the Hilbert space is simply given by the angular momentum states $|j m\rangle$. Hence, the naive transition amplitude due to the above may be computed as follows
\begin{align}
   W(\mathcal{M}) = \operatorname{Tr}\left(\prod_{x_i\in\mathcal{M}}e^{-i H_{x_i}}\right) \label{TAmp1}
\end{align}

where the trace here includes the trace over the angular momentum states and an integration over the connection 1-form which ends up becoming the Haar measure for $SU(2)$. This is very similar to the transition amplitude computed in LQG\cite{rovelli2015covariant}. To do so, we insert the following complete set of states\cite{shankar2012principles, napolitanomodern} 
\begin{align}
    \sum_{j,m}(2j+1)|j m\rangle \langle j m| = I
\end{align}
When inserted between every exponential, so that at every point on the manifold, we have
\begin{align}
    \int [\mathcal{D}\mathcal{\omega}]  \langle j_{i}m_i| e^{\frac{i}{2}\omega_A\wedge J^A}|j_{i} n_{i}\rangle = \int [\mathcal{D}\mathcal{\omega}] D^{(j_i)}_{m_in_i}(\omega) = \frac{8\pi^2}{2j_i+1}\delta_{m_i, 0}\delta_{n_i, 0}\delta_{j_i, 0}
\end{align}  
where as indicated before $[D\omega]$ the Haar measure for $SU(2)$. Due to which the partition function simply becomes
\begin{align}
    W(\mathcal{M})=\sum_{j_x}\prod_{x \in \mathcal{M}}1
\end{align}

This is quite anti-climactic. However, this is merely a demonstration of the non-quantizable nature of differential geometry, more specifically, pseudo-Riemannian geometry. In other words, pseudo-Riemannian manifolds resist attempts at direct quantization. That is why, it is much easier to quantize an alternative geometry from which Riemannian manifolds emerge as a limiting or effective geometry as is done in LQG\footnote{In LQG, the discrete version of the EH action called the Regge action, which is essentially a theory of Regge manifolds, is quantized. The Regge geometry gives the usual differential geometry as a limiting case. The Regge manifold as a consequence is also the quantum mechanical description of the spacetime geometry.} or string theory\cite{polchinski2005string, Tong:2009np}, respectively.
\subsection{Quantizable geometry: An illustration}
\begin{figure}[h!]
    \centering
    \includegraphics[scale=0.20]{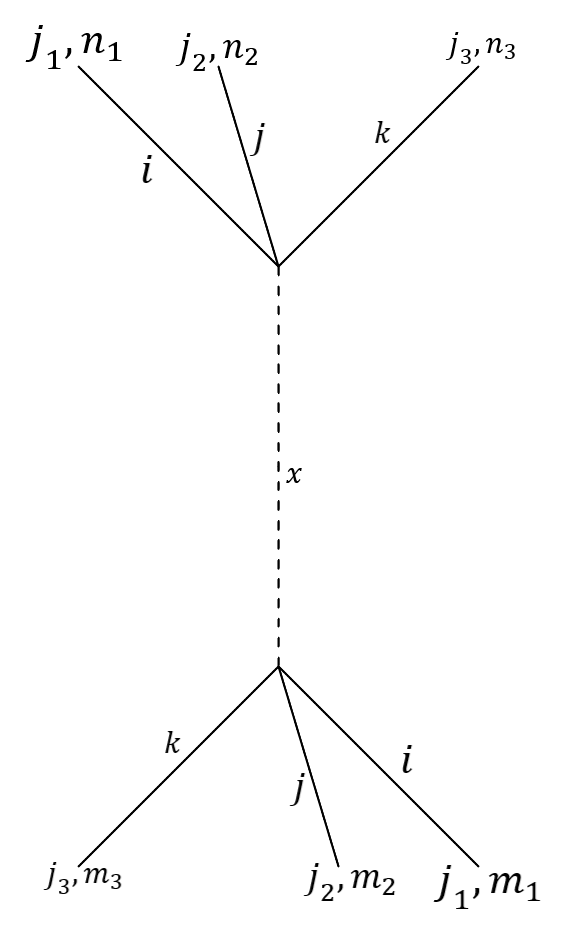}
    \caption{The dotted line represents the position $x$ on the manifold. The bold lines represent the momentum label e.g. a line with the end-label $J_1$ and mid-label letter $i$ represents the momentum label $(j_{1i}, m_{1i})$. The two endpoints of the dotted-line represents the two $3j$ symbol at each position label $x$.}
    \label{fig:int}
\end{figure}

\begin{figure}[h!]
    \centering
    \includegraphics[scale=0.40]{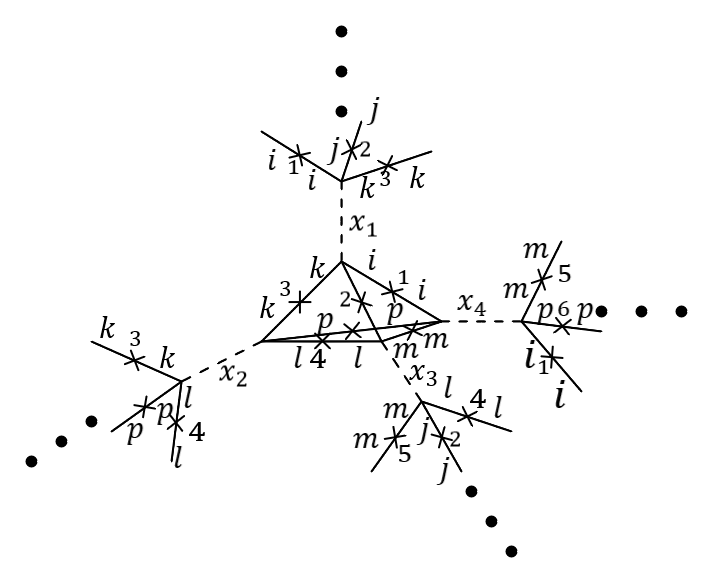}
    \caption{The cross represents the contraction of the indicies labeled by the letters on the either side of the cross and a number on the cross e.g. a crossed line with number $4$ and letter $l$ represents a contraction either the index $m_{4l}$ or $n_{4l}$ corresponding to the momenta $j_{4l}$. $m$ and $n$ labels are contracted only with other $m$'s and $n$'s, respectively. A tetrahedral contraction like the above is the $6j$-symbol. The bold dots represent continuity of similar tetrahedral contractions throughout the manifold.}
    \label{fig:cont}
\end{figure}

Since, usual pseudo-Riemannian geometry is not `quantizable', let us then move on to the non-Riemannian sector of geometry. To make the underlying geometry non-Riemnannian, we include the torsion term\footnote{More specifically Riemann-Cartan manifold.} to the action as follows
\begin{align}
    S  = \frac{1}{2\pi G_N}\int_{\mathcal{M}}(e_A+\eta D\lambda_A) \wedge F^{A} +\gamma\int_{\mathcal{M}}e_A\wedge De^A\label{torterm}
\end{align}
along with a 0-form $\lambda^A$. The quantization procedure is followed as before. The conjugate momenta for the modified action are then given by
\begin{align}
    &\pi_{A}^{\omega} = \frac{\delta S}{\delta (d\omega^{A})} =\frac{e_A+\eta D\lambda_A}{2\pi G_N}
    \\
    & \pi^e_A = \frac{\delta S}{\delta (de^A)} = \gamma e_A\\
    & \pi^{\lambda}_A = \frac{\delta S}{\delta (d\lambda^A)} =  \frac{\eta}{2\pi G_N}F_A
\end{align}
with the following Poisson brackets
\begin{align}
    \{\pi^{\omega}_A, \omega^{B}\} = \delta_A^{~B} \quad \{\pi^e_{A}, e^B\} = \delta_A^B \quad \{\pi^\lambda_A, \lambda^B\} =   \delta_A^{~B}
\end{align}
where the first two are anticommuting  brackets while the last one is a commuting bracket.
Then we can identify the following primary constraint
\begin{align}
\chi_{A}\equiv\pi_{A}^{e} -\gamma e_A\approx 0 
\end{align}
The constraint is second class. Then the preliminary Hamiltonian density is as follows
\begin{align}
    H_1 =&-\frac{1}{2}\epsilon^{ABC}\pi^{\omega}_A\wedge\omega_{B}\wedge \omega_{C}-\epsilon^{ABC}\pi^e_A\wedge\omega_B\wedge e_C-\eta\epsilon^{ABC}\pi^{\lambda}_{A}\wedge\omega_B\wedge\lambda_C\notag\\
    &+\pi_{\lambda}^A\wedge\left(2\pi G_N \pi^{\omega}_A-\frac{\pi^e_A}{\gamma}\right) + c^A\wedge \chi_A
\end{align}
The Dirac matrix is simply
\begin{align}
    \mathcal{C}_{AB} = -2\gamma\delta_{AB} \quad  \mathcal{C}^{-1}_{AB} = -\frac{\delta_{AB}}{2\gamma}
\end{align}
which lead to the following Dirac brackets
\begin{align}
    &\{\pi^e_A, e^B\}_{D} = \frac{\delta_A^B}{2} \quad  \{e^A, e^B\}_{D} = \frac{\delta^{AB}}{2\gamma} \quad \{\pi^e_A, \pi^e_B\}_{D} = \frac{\gamma\delta_{AB}}{2}\\
    &\{\pi^{\omega}_A, \omega^B\}_{D} = \delta_A^B \quad \{\pi^{\lambda}_A, \lambda^B\}_{D} = \delta_A^B
\end{align}
We do the following redefinition of the canonical variables
\begin{align}
    \tilde{\pi}^e_A \equiv \sqrt{\frac{2}{\gamma}} \pi^e_A\quad \tilde{e}^A \equiv \sqrt{2\gamma} e^A
\end{align}
We set $\eta = \frac{1}{2}$ and as an approximation take $\gamma \to \infty$ keeping $2\pi G_N\sqrt{\gamma}$ fixed so that the Dirac Hamiltonian looks like
\begin{align}
    H \approx&-\frac{1}{2}\epsilon^{ABC}\pi^{\omega}_A\wedge\omega_{B}\wedge \omega_{C}-\frac{1}{2}\epsilon^{ABC}\tilde{\pi}^e_A\wedge\omega_B\wedge \tilde{e}_C-\frac{1}{2}\epsilon^{ABC}\pi^{\lambda}_{A}\wedge\omega_B\wedge\lambda_C
\end{align}
The Dirac brackets are raised to the quantum oscillator algebra
\begin{align}
    &[\pi^{\omega}_A, \omega^{B}]_{+} = i\delta_A^{~B} \quad [\pi^e_{A}, e^B]_{+} = i\delta_A^B \quad [\pi^{\lambda}_{A}, \lambda^B]_{-} = i\delta_A^B \\
    & [\tilde{\pi}^e_A, \tilde{e}^B]_{+} = i\delta_A^B \quad  [\tilde{e}^A, \tilde{e}^B]_{+} = i\delta^{AB}\quad [\tilde{\pi}^e_A, \tilde{\pi}^e_B]_{+} = i\delta_{AB}
\end{align}
Just like in the previous section, we can define
\begin{align}
    &J^A_1 = \epsilon^{ABC}\pi^{\omega}_B\wedge\omega_C \quad J^A_2 = \frac{1}{2}\epsilon^{ABC}\tilde{\pi}^{e}_B\wedge \tilde{e}_C \quad J^A_3 =  \epsilon^{ABC}\pi^{\lambda}_B\wedge\lambda_C
\end{align}
All the three satisfy 
\begin{align}
    &[J_i^A, J_i^B]_{-} = i\epsilon^{ABC}J_i^C \quad [J_i, J_j]_{-} =  0 \quad i\neq j\quad 
    \forall i, j = 1, 2, 3
\end{align}
So that the Hamiltonian becomes
\begin{align}
H \approx -\frac{1}{2}\omega_A\wedge (J^A_1+2J^A_2+J^A_3)  \equiv -\frac{1}{2}\omega_A\wedge \mu^A \label{DHam}
\end{align}
Notice that $\mu^A$ is a vector operator as
\begin{align}
    [\mu^A, \mathcal{J}^B]_{-} = i\epsilon^{ABC}\mu_C \quad \mathcal{J}^A = J^A_1+J^A_2+J^A_3
\end{align}
Since, the Hamiltonian contains an $SU(2)$ vector operator $\mu^A$, hence, the Hilbert space is again spanned by the angular momentum states $|j, m\rangle$. Now, we wish to compute
\begin{align}
    W(\mathcal{M}) = \operatorname{Tr}\left(\prod_{x\in\mathcal{M}}e^{-i H_x}\right)\label{PI}
\end{align}
where the trace is the same as the one used in the transition amplitude \ref{TAmp1}. To do so, we insert the following complete set of states 
\begin{align}
    \sum_{j,m}(2j+1)|j m\rangle \langle j m| = I
\end{align}
When inserted between every exponential, then at every point we have
\begin{align}
    \int [\mathcal{D}\mathcal{\omega}]  \langle j_{1i} m_{1i}~ j_{2j} m_{2j}~ j_{3k} m_{3k}| e^{\frac{i}{2}\omega_A\wedge\mu^A}|j_{1i} n_{1i}~ j_{2j} n_{2j}~ j_{3k} n_{3k}\rangle
\end{align}
where
\begin{align}
|j_1 m_1 ~j_2 m_2 ~j_3 m_3\rangle \equiv |j_{1} m_{1}\rangle \otimes |j_{2} m_{2}\rangle \otimes |j_{3} m_{3}\rangle
\end{align}
To evaluate the above, we make use of the Wigner-Eckart projection theorem\footnote{See Appendix \ref{A}}, to write the above as\cite{rovelli2015covariant}
\begin{align}
    &\int [\mathcal{D}\mathcal{\omega}]  \langle j_{1i} m_{1i}~ j_{2i} m_{2i}~ j_{3i} m_{3i}| e^{\frac{i}{2}\omega_A\wedge\mu^A}|j_{1i} n_{1i}~ j_{2j} n_{2j}~ j_{3k} n_{3k}\rangle\notag\\
    &=\int [\mathcal{D}\mathcal{\omega}]  \langle j_{1i} m_{1i}~ j_{2i} m_{2i}~ j_{3i} m_{3i}| e^{f_{j_{1i}, j_{2j}, j_{3k}}\frac{i}{2}\omega_A\wedge\mathcal{J}^A}|j_{1i} n_{1i}~ j_{2j} n_{2j}~ j_{3k} n_{3k}\rangle\notag\\
    &=\frac{1}{f_{j_{1i}, j_{2j}, j_{3k}}^3}\int [\mathcal{D}\mathcal{\omega}] D^{(j_{1i})}_{m_{1i}n_{1i}}(\omega) D^{(j_{2j})}_{m_{2j}n_{2j}}(\omega) D^{(j_{3k})}_{m_{3k}n_{3k}}(\omega)=\frac{\i^{m_{1i}m_{2j}m_{3k}}\i^{n_{1i}n_{2j}n_{3k}}}{f_{j_{1i}, j_{2j}, j_{3k}}^3}\notag\\
    &=\frac{1}{f_{j_{1i}, j_{2j}, j_{3k}}^3}\begin{pmatrix}
        j_{1i} && j_{2j} && j_{3k}\\
        m_{1i} && m_{2j} && m_{3k}
    \end{pmatrix}
    \begin{pmatrix}
        j_{1i} && j_{2j} && j_{3k}\\
        n_{1i} && n_{2j} && n_{3k}
    \end{pmatrix}\label{3jsymb}
\end{align}
where
\begin{align}
    f_{j_{1i}, j_{2j}, j_{3k}} &= \frac{\langle j_{1i} m_{1i}~ j_{2j} m_{2j}~ j_{3k} m_{3k}|\mu^{A}\wedge\mathcal{J}_A|j_{1i} m_{1i}~ j_{2j} m_{2j}~ j_{3k} m_{3k}\rangle}{\langle j_{1i} m_{1i}~ j_{2j} m_{2j}~ j_{3k} m_{3k}|\mathcal{J}^2|j_{1i} m_{1i}~ j_{2j} m_{2j}~ j_{3k} m_{3k}\rangle}=\frac{3}{2}+\frac{\mathcal{J}_2(\mathcal{J}_2+1)-\mathcal{J}_{13}(\mathcal{J}_{13}+1)}{2\mathcal{J}(\mathcal{J}+1)}
\end{align}
Therefore, once the integral over the 1-forms are performed at every point, we get a bunch of $3j$-symbols that needs to be contracted amongst each other. Diagrammatically, we can represent the above integral as Figure \ref{fig:int}. It is helpful to think of it as a Feynmann interaction vertex in QFT.
Now, just like in the QFT transition amplitudes, we only focus on the `connected diagrams'. In this case, Figure \ref{fig:cont} represents the analogue of 1PI irreducible connected diagrams of QFT.

Figure \ref{fig:cont} is very much reminiscent of the dual of the triangulation of a $3D$ manifold that makes an appearance in the computation of the transition amplitude in LQG\cite{rovelli2015covariant}. Therefore, after all the contractions in this 1PI connected diagram, we obtain
\begin{align}
    [W(\mathcal{M})]_c = \sum_{j_{1x_i}, j_{2x_i}, j_{3x_i}}\prod_{x_i \in \mathcal{M}} \frac{(2j_{1x_i}+1)(2j_{2x_i}+1)(2j_{3x_i}+1)}{f^3_{j_{1x_i}, j_{2x_i}, j_{3x_i}}}\prod_{(x_a, x_b, x_c, x_d)}(-1)^{j_{abcd}}\{6j\}
\end{align}
where the subscript $c$ represents the connected part of the transition amplitude and $(x_a, x_b, x_c, x_d)$ represents the four points in the manifold where the tetrahedral contraction of the $3j$-symbols happen to give the $\{6j\}$ symbol. Different 1PI contractions are analogous to the different triangulations of the $3D$ manifold that one can do. If in Figure \ref{fig:cont}, we call all crossed lines as $e$ and the tetrahedrons as $v$ and redefine the intertwiner in \ref{3jsymb} as
\begin{align}
    \tilde{\i}^{m_{1i}m_{2j}m_{3k}} \equiv \frac{\i^{m_{1i}m_{2j}m_{3k}}}{f^{3/2}_{j_{1i}, j_{2j}, j_{3k}}}
\end{align}
and redefine the corresponding tetrahedral contractions of the intertwiners as $\{\widetilde{6j}\}$, then we can rewrite the above transition amplitude in a more familiar form as follows
\begin{align}
    [W(\mathcal{M})]_c = \sum_{j_e}\prod_{e} (2j_e+1)\prod_{v}(-1)^{j_v}\{\widetilde{6j}\}
\end{align}
Except for the annoying scaling factors of $f^{3/2}_{j_{1i}, j_{2j}, j_{3k}}$ that sits within these symbols, the answer bears a very close resemblance to the Ponzano-Regge amplitude of LQG. If we pay a close attention to the modified action we can see that
\begin{align}
    D\lambda_A\wedge F^A = D(\lambda_A\wedge F^A) = d(\lambda^A\wedge F^A)
\end{align}
which is an exact form due to the Bianchi identity\cite{nakahara2018geometry} and, hence, doesn't play any role in the bulk of the manifold. Hence, the only relevant addition in the modified action is the torsion term corresponding to the coupling $\gamma$. Therefore, by adding a torsion term, we have made the manifold `quantizable'. The addition of the torsion term `enforces' a triangulation on an a priori smooth, differentiable manifold. This is exactly the opposite of LQG where the triangulation of a manifold is quantized first and smooth, differentiable geometry emerges as a limiting case. Now, one may say that this modification of the geometry is a bit `extreme'. Perhaps, there are less extreme modifications of the geometry that can reproduce the Ponzano-Regge amplitude of LQG exactly? Unfortunately, despite our best efforts we have not been able to find anything simpler than \ref{torterm} that can make the manifold exactly `quantizable'\footnote{As in, get rid of the $f_{j_{1i}, j_{2j}, j_{3k}}$ factor.}.
However, this exercise very well illustrates the meaning of `quantizable' geometry in the manner we intend to define, which is 
\begin{definition}[Quantizable geometry]
A mainfold $\mathcal{M}$ is said to be quantizable if it allows for a self-consistent Dirac Quantization procedure on $\oplus_{r}\Omega^r(\mathcal{M})$ such that the following map 
\begin{align}
    \Omega^{1}(\mathcal{M}) \overset{Dirac}{\rightarrow} \{\theta\} 
\end{align}\label{Defn}
holds.
\end{definition}
\newpage
\subsection{Quantizing triangulations vs Quantizing manifolds}
\begin{figure}[h!]
    \centering
    \includegraphics[scale=0.4]{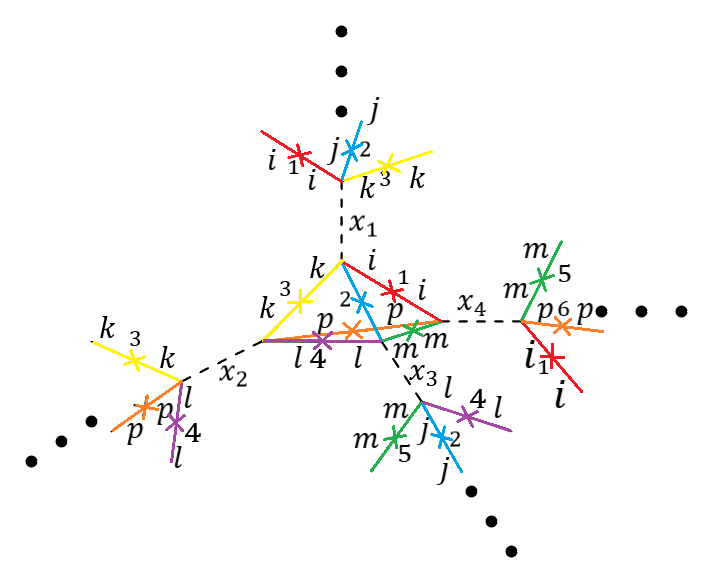}
    \caption{The lines with same colours are labeled by the same $j_i$. Eg. Red: $j_1$ Blue: $j_2$ Yellow: $j_3$ Purple: $j_4$ and so on...}
    \label{fig:cont:cc}
\end{figure}

Let us assume for a moment that there exists a geometry that is exactly `quantizable' i.e. in \ref{DHam} we instead have $\mu \to \mathcal{J}$ which then implies $f_{j_{1i}, j_{2j}, j_{3k}} \to 1$. Then it is easy to see that we will have
\begin{align}
    [W(\mathcal{M})]_{c} = \sum_{j_e}\prod_{e} (2j_e+1)\prod_{v}(-1)^{j_v}\{6j\}
\end{align}
which seems to match the Ponzano-Regge amplitude exactly. However, if one takes a closer look at the contractions that take place in Figure \ref{fig:cont}, one can notice that the various angular momenta label $j_i$ are repeated infinitely throughout the manifold. To make this clear, we colourcoded Figure \ref{fig:cont} in Figure \ref{fig:cont:cc} to highlight the repeating labels throughout the contractions. \\\\
This is unlike the Ponzano-Regge amplitudes where the $j$ labels are arbitrary for each edge of the dual of the triangulation. This is a demonstration of the subtle difference that can arise between quantizing a manifold and quantizing a triangulation. These repeating $j$-labels can be thought of as crack lines that appear in a glass. If one thinks of differential manifold as a glass object, attempts to `quantize' differential forms metaphorically `cracks' the spacetime at the quantum level. LQG on the other hand is like `glassmaking' a differential manifold by joining various chunks of `glass'\footnote{Read spacetime.} at the quantum level seamlessly without `cracks'. Different 1PI contractions of the kind in Figure \ref{fig:cont:cc} are, therefore, analogous to the different ways the same glass object can be cracked. So, even if there exists a `quantizable' geometry, such a geometry will exhibit such subtle characteristic differences from the established methods of quantization such as the LQG.\\
\section{Discussion}
In summary, an attempt was made to quantize $3D$ pseudo-Riemannian manifolds by directly `quantizing' differential forms. The exterior derivative $d$ was interpreted as the `time' derivative for the differential forms. This interpretation was shown to consistently reproduce the Euler-Lagrange equation as was demonstrated with the example of the scalar field action. Emboldened by the seeming efficacy of this interpretation, it was finally employed to compute the momenta conjugate to connection and triad 1-form in $3D$ Einstein-Hilbert action and initiate the Dirac Quantization procedure for the same. Initial quantization attempt did not yield any significant result. Until the manifold in question was made non-Riemannian by adding torsion terms to the EH action where this time the quantization yielded a fruitful result i.e. an object that very closely resembles the transition amplitude obtained via LQG. This is interesting because even though transition amplitudes have been computed before using the Dirac Quantization procedure for $2+1$-gravity in metric formalism\cite{yamada1990transition}, it is not immediately clear how those results are related to the results of LQG. But with the Dirac procedure as proposed in the paper, we can replicate the results of LQG very closely.\\

Through this exercise, the main question we tried to answer was, if it is possible to understand the quantum mechanical nature of the spacetime geometry by implementing the standard Dirac Quantization procedure to the EH action in the first-order formalism. Or to put it simply, can Dirac Quantization of gravity help us `derive' string theory, LQG or any other well-known theory of quantum gravity? This is completely opposite to the approach that is usually taken in formulating a quantum theory of gravity. In this exercise, we learnt that pseudo-Riemannian geometry is not quantizable while non-Riemannian geometry is `quantizable' to the extent that it can at least qualitatively\footnote{Perhpas, even approximately.} capture the quantum mechanical nature of spacetime geometry as predicted by the LQG.\\

However, as was demonstrated later on, even if we can describe the GR action via a truly `quantizable' geometry, the observables computed would still slightly differ from the observables computed via LQG. This is where the analogy of differential manifolds to glass objects becomes uncanny. The quantization procedure implemented in this paper is akin to deliberately inducing cracks in the glass compared to seamlessly `glassmaking' the differential manifold from `quantized' chunks of glass like in the LQG. This precisely answers the question posed a paragraph ago. Dirac Quantization of differential geometry, in the way proposed in this paper, may help us glimpse the quantum mechanical nature of spacetime geometry consistent with well-established formulations of quantum gravity.\\

Of course, one may ask if there is a way where one can achieve this exactly rather than just qualitatively or approximately. Maybe, the exterior derivative $d$ is not exactly the right `time' derivative. There are, perhaps, other well-defined operations on $\oplus_{r}\Omega^r(\mathcal{M})$ that can work as a `time' derivative. Or, perhaps, the differential forms are the not exactly the right objects that must be quantized to quantize differential geometry. These are the possibilities which we leave for our future investigations. However, the more important issue is whether this quantization procedure can be extended to spacetimes of other dimensions. This is where we must contend with a depressing reality that this quantization procedure is only special to $3D$ manifolds. To understand this, let's look at the anticommuting Poisson brackets for 1-forms in \ref{APB}. One can notice that it is a matter of great fortune that Possion brackets can be defined so consistently for $3D$ manifolds this way. To see this, consider the, naive, operator form of the conjugate momenta as follows
\begin{align}
    \pi_A^{\omega} \to -i\frac{\delta}{\delta\omega^A} \quad   \pi_A^{e} \to -i\frac{\delta}{\delta e^A} 
\end{align}
Now because the triad and connection 1-forms have ``grassmannian-odd'' statistics, therefore, their respective variational derivatives must carry the same statistics. This is consistent with the statistics of the primary constraints obtained in \ref{pc}. However, this fortune is not endowed to us if one looks at the $4D$ EH action in first-order formalism
\begin{align}
    S = \frac{1}{2\pi G_N}\int_{\mathcal{M}} e^A\wedge e^B \wedge \star F_{AB}
\end{align}
The primary constraints here look like
\begin{align}
    &\pi_{\omega}^{AB}-\frac{\star e^A\wedge e^B}{2\pi G_N} \approx 0\\
    &\pi^e_A \approx 0
\end{align}
while the naive operator form for these conjugate momenta should be
\begin{align}
    \pi^{AB}_{\omega} \to -i\frac{\delta}{\delta\omega_{AB}} \quad   \pi_A^{e} \to -i\frac{\delta}{\delta e^A} 
\end{align}
which carry the ``grassmannian-odd'' statistics same as the $3D$ case, which is in stark contrast to the primary constraint above where $\star e^A\wedge e^B$ has ``grassmannian-even'' statistics. This means that the conjugate momenta and Poisson brackets for connection and tetrad 1-forms cannot be defined consistently for $4D$ Riemannian manifolds, therefore, violating the criteria of Definition \ref{Defn} for `quantizable' manifolds. Unlike the $3D$ case, where the criteria of the Definition \ref{Defn} was at least satisfied superficially. So, it may very well be that $3D$ gravity is indeed special and the fact that we were able to get this far is nothing short of a miracle. Having said that, there may be some unexplored alternatives or techniques that can help us extend the Dirac Quantization procedure for differential forms to arbitrary spacetime dimensions. Or, perhaps, one may need to quantize some exotic geometric structures instead. This is again something we intend to look for in our future investigations.\\
\section*{Acknowledgements}
I would like to acknowledge the support and kindness of my supervisor and employer Prof. Junggi Yoon (JRG, Holography and Black Holes), APCTP, Pohang, Republic of Korea without which this initiative would not have been possible. APCTP is supported through the Science and Technology Promotion Fund and Lottery Fund of the Korean Government. I would also like to acknowledge the moral support of Hare Krishna Movement, Pune, India. I would also like to thank Carlo Rovelli for answering many of my questions in LQG over mail. This research is dedicated to the people of India and the Republic of Korea for their steady support of research in theoretical science. This work was supported by the NRF grant funded by the Korea government (MSIT) (No. 2022R1A2C1003182).\\
\appendix 
\renewcommand{\theequation}{A.\arabic{equation}}
\setcounter{equation}{0}
\section{Wigner-Eckart theorem for exponentiated vector operators}\label{A}
Consider a vector operator $\mu^{A}$ such that it satisfies
\begin{align}
    [\mu^A, \mathcal{J}^B] = i\epsilon^{ABC}\mu^C \quad [\mathcal{J}^A, \mathcal{J}^B] = i\epsilon^{ABC}\mathcal{J}^C
\end{align}
Now, consider the following object
\begin{align}
    \langle j m|e^{i\theta\cdot\mu}|j m'\rangle &= \sum_{n = 1}^{\infty}i^n\frac{\langle j m|(\theta\cdot\mu)^n|j m'\rangle}{n!} \notag\\&= \sum_{n = 1}^{\infty}\frac{i^n(2j+1)^{n-1}}{n!}\sum_{m_i = -j}^j\langle jm|\theta\cdot\mu|jm_n\rangle\sum^{n-2}_{i = 1}\langle j m_{i+1}|\theta\cdot\mu|j m_i\rangle\langle jm_1|\theta\cdot\mu|jm'\rangle
\end{align}
where in the above the complete set of states
\begin{align}
    \sum_{j,m}(2j+1)|j m\rangle \langle j m| = I
\end{align}
was inserted. Now, we make use of the Wigner-Eckart projection theorem given by\cite{napolitanomodern, shankar2012principles}
\begin{align}
    \langle jm|\mu^A|jm'\rangle = \frac{\langle jm| \mu\cdot\mathcal{J}|jm'\rangle}{\langle jm|\mathcal{J}^2|jm'\rangle}\langle jm|\mathcal{J}^A|jm'\rangle \equiv \frac{\langle \mu\cdot\mathcal{J}\rangle}{\langle \mathcal{J}^2\rangle}\langle jm|\mathcal{J}^A|jm'\rangle 
\end{align}
where $\langle \mu\cdot\mathcal{J}\rangle/\langle \mathcal{J}^2\rangle$ is a coefficient independent of $m, m'$. We make use of this in the above to get
\begin{align}
    \langle j m|e^{i\theta\cdot\mu}|j m'\rangle = \sum_{n = 1}^{\infty}\frac{i^n\langle\mu\cdot\mathcal{J}\rangle^n}{\mathcal{J}^n(\mathcal{J}+1)^nn!}\langle jm|(\theta\cdot\mathcal{J})^n|jm'\rangle = \langle jm|\exp\left(\frac{\langle\mu\cdot\mathcal{J}\rangle}{\mathcal{J}(\mathcal{J}+1)}\theta\cdot\mathcal{J}\right)|jm'\rangle 
\end{align}
\bibliographystyle{plain} 
\bibliography{refs}
\end{document}